\documentstyle[multicol,eqsecnum,aps]{revtex}
\renewcommand{\narrowtext}{\begin{multicols}{2}
\global\columnwidth20.5pc\noindent}
\renewcommand{\widetext}{\end{multicols}
\global\columnwidth42.5pc}
\begin{document}
\draft

\title{\bf Optical conductivity of the Kondo 
insulator YbB$_{12}$: \\
Gap formation and low-energy excitations}
\author{H. Okamura}
\address{Department of Physics, Kobe University, 
Kobe 657-8501, Japan.}
\author{S. Kimura}
\address{Graduate School of Science and Technology, 
Kobe University, Kobe 657-8501, Kobe, Japan, \\
and UVSOR Facility, Institute for Molecular Science, 
Okazaki 444-8585, Japan}
\author{H. Shinozaki and T. Nanba}
\address{Graduate School of Science and Technology, 
Kobe University, Kobe 657-8501, Kobe, Japan}
\author{F. Iga, N. Shimizu, and T. Takabatake} 
\address{Graduate School of Advanced Science of Matter, 
Hiroshima University,\\ Higashi-Hiroshima 739-8526, Japan.}
\maketitle
\begin{abstract}
Optical reflectivity experiments have been conducted 
on single crystals of the Kondo insulator YbB$_{12}$ 
in order to obtain its optical conductivity, 
$\sigma(\omega)$.    Upon cooling below 70~K, a strong 
supression of $\sigma(\omega)$ is seen in the far-infrared 
region, indicating the opening of an energy gap of $\sim$~25~meV.    
This gap development is coincident with a rapid decrease 
in the magnetic susceptibility, which shows that the gap 
opening has significant influence on magnetic properties.   
A narrow, asymmetric peak is observed at $\sim$~40~meV in 
$\sigma(\omega)$, which is attributed to optical transitions 
between the Yb $4f$-derived states across the gap.   
In addition, a broad peak is observed at $\sim$~0.25~eV.   
This peak is attributed to transitions between Yb 
$4f$-derived states and $p$-$d$ band, and is reminiscent 
of similar peaks previously observed for rare-earth 
hexaborides.  
\end{abstract}

\narrowtext

The term ``Kondo insulators'', or equivalently ``Kondo 
semiconductors'', refers to a group of strongly 
correlated $f$-electron compounds that exhibit the following 
characteristic behaviors.   At high temperatures they 
behave as Kondo lattice systems with local magnetic 
moments and are often metallic.   In contrast, 
at low temperatures they become semiconducting as a small 
energy gap develops at the Fermi energy ($E_F$), 
and their magnetic susceptibility becomes much 
smaller.\cite{fisk,takabatake}     
While their behaviors have been modeled qualitatively 
by a hybridization of $f$- and conduction bands 
renormalized by correlation effects\cite{fisk}, the 
microscopic mechanism for the gap formation is still unclear.  
In particular, it is not well understood how 
the local Kondo coupling plays a role in the gap formation.  
To address these questions, it is important to obtain 
detailed information about low energy excitations 
near $E_F$.     Optical reflectivity studies have been 
a useful tool for this purpose, and have provided 
much information on Kondo insulators such as 
SmB$_6$ [Ref.~\onlinecite{wachter1,nanba1,kimura1}] and 
Ce$_3$Bi$_4$Pt$_3$ [Ref.~\onlinecite{ce343}].  

Among the Kondo insulators, YbB$_{12}$ is the only known 
Yb-based compound, and it has been studied 
extensively.\cite{yb12-old,wachter2,susaki}    The magnetic 
susceptibility of YbB$_{12}$ shows a Curie-Weiss behavior 
at high temperatures, but it decreases rapidly 
upon cooling below $\sim$~70~K.     
The presence of an energy gap has been shown by 
activation-type temperature dependences in dc resistivity, 
specific heat, and Hall effect measurements, with measured 
gap $2\Delta$ in the range 10-15 meV.\cite{yb12-old}  
Previously, only sintered, polycrystalline samples were 
available due to difficulties in crystal growth, 
which limited a reliable optical 
measurement to the far-infrared.\cite{wachter2}   
Recently, however, Iga {\it et al.}\cite{yb12-single} 
have successfully grown large single crystals of YbB$_{12}$ and 
other rare-earth dodecaborides.  

In this work, we report the first optical reflectivity 
measurements of single crystals YbB$_{12}$, at photon 
energies between 0.008 eV and 50 eV and at temperatures 
between 20 K and room temperature.   The resulting optical 
conductivity spectrum, $\sigma(\omega)$, has revealed many 
optical transitions that were previously unknown.    
Below 70~K, 
$\sigma(\omega)$ clearly shows the formation of an energy 
gap of $\sim$~25~meV.    This gap opening is coincident with 
a rapid decrease in the magnetic susceptibility, showing that 
the gap opening has large effects on the magnetic 
properties.   Other low-energy excitations are analyzed 
in terms of the $f$- and the $pd$-derived states near $E_F$.   
We have also studied single crystals of LuB$_{12}$, 
which is a non-magnetic metal with a filled 4$f$ shell, for 
comparison to YbB$_{12}$.

Single crystals of YbB$_{12}$ and LuB$_{12}$ 
were grown by the floating zone method using a newly 
developed image furnace equipped with four Xe 
lamps\cite{yb12-single}.  Both compounds have a cubic, 
NaCl-type structure consisting of Yb or Lu ions and 
B$_{12}$ cubo-octahedrons.    The samples, with 
typical dimensions of $\sim$~6~ mm diameter 
and $\sim$~1~mm thickness, were polished and 
mounted on a liquid He cryostat, which was inserted 
into a vacuum chamber that contained the necessary optics 
for reflectivity measurements under near-normal incidence.   
An evaporated Au or Al film was mounted next to the samples 
as a reference mirror.    For measurements below 2.5 eV, 
a rapid-scan Fourier interferometer (Bruker Inc. IFS-66v) 
was used with various combinations of beam splitters, 
detectors and light sources, including synchrotron 
radiation at beamline BL6A1 of the UVSOR Facility, 
Institute for Molecular Science.\cite{bl6a1}   
Measurements between 2 and 50 eV were performed 
at room temperature only, using synchrotron radiation 
at beamlines BL1B and BL5B of UVSOR.   The reflectivity 
of the samples was also checked using He-Ne and Ar lasers 
at several photon energies.  Standard Kramers-Kronig analyses 
were made to obtain $\sigma(\omega)$ from a measured 
reflectivity spectrum $R(\omega)$, combined with a Hagen-Rubens 
extrapolation [$R(\omega) \propto 1-a\sqrt{\omega}$]
to the lower-energy end and a $R(\omega) \propto~\omega^{-4}$ 
extrapolation for the higher-energy end\cite{footnote3}.  

Figure 1 shows $R(\omega$) and $\sigma(\omega)$ of 
YbB$_{12}$ and LuB$_{12}$ single crystals 
at room temperature.   For both compounds, 
a sharp onset is seen in $R(\omega)$ near 1.6 eV, 
which can be identified as the plasma edge ($\omega_p$) 
due to a metallic response of mobile carriers.   The peak 
structures above $\omega_p$ are due to interband 
transitions between electronic states far apart from 
$E_F$.\cite{kimura2}       The similarity of these spectra 
above $\omega_p$ for YbB$_{12}$ and LuB$_{12}$ shows 
that these states are nearly unaffected by Yb/Lu replacement.   
Below $\omega_p$, in contrast, 
the spectra are strikingly different for the two compounds: 
In LuB$_{12}$, $R(\omega)$ is nearly flat and 
$\sigma(\omega)$ shows a sharp rise, 
which is typical of a good metal, while in YbB$_{12}$ 
there is a broad dip in $R(\omega)$ giving 
rise to a strong peak at 0.25 eV in $\sigma(\omega)$.   
Hereafter we refer to this peak as the ``IR peak'', 
and concentrate on the spectra below $\omega_p$.      
Detailed analyses on the interband transition peaks 
above $\omega_p$ will be presented 
elsewhere.\cite{kimura3}   

Figure 2 shows $R(\omega)$ and $\sigma(\omega)$ 
for YbB$_{12}$ measured at 290, 160, 78, and 20 K.   
As the temperature is lowered from 290 K to 78 K, 
$\sigma(\omega)$ is gradually reduced over the entire 
infrared region, and the IR peak becomes enhanced 
and slightly blue-shifted.  At 78 K the spectra are 
still metallic, in the sense that $R(\omega)$ 
approaches 1 at the lower energy end.  At 20 K, 
however, the spectral weight below $\sim$~40~meV 
in $\sigma(\omega)$ is strongly depleted; 
the spectrum is now typical of an insulator 
(semiconductor) with an energy gap.     
Note that, on cooling from 
78 K to 20 K, the spectral weight lost by the gap 
formation in $\sigma(\omega)$ is transferred to 
the higher-energy side of the IR peak.    
Namely, the optical sum rule is 
satisfied by transfers of spectral weight over an 
energy scale of $\sim$~1~eV, rather than by 
transfers to directly above the gap.   
This point will be discussed later.  
The temperature dependence of $R(\omega)$ and 
$\sigma(\omega)$ below 78~K is 
shown in Fig.~3 for the low energy region.   
These spectra demonstrate that the gap develops 
progressively over the temperature range $T \leq 70$~K.   
At 20~K the gap 
appears fully developed, with an ``onset'' of 
$\sigma(\omega)$ at $\sim$~25~meV.  
Above the onset, $\sigma(\omega)$ rises quickly to 
a ``shoulder'' at $\sim$~40~meV, which is marked 
by the arrow in Fig.~3.   Since the specral 
weight below the onset is very small, we identify 
the onset energy as the magnitude of the gap, 
$2\Delta_{opt} \simeq$ 25~meV.    

There are remarkable aspects in the 
temperature dependence and the magnitude of the 
observed gap in $\sigma(\omega)$.   
The inset of Fig.~3 shows $\chi(T)$, the magnetic 
susceptibility of the single crystal YbB$_{12}$ 
as a function of temperature.\cite{yb12-single}    
$\chi(T)$ has a broad maximum at $T_{max} \sim 75$~K.   
Below $T_{max}$ it decreases rapidly with cooling, 
and results in a finite value at low 
temperatures.\cite{takabatake,footnote}    
Comparing the optical spectra in Fig.~3 
with $\chi(T)$, it is clear that 
the gap development coincides with the rapid 
decrease of $\chi(T)$ in exactly the 
same temperature range.      
Although it is very likely that the gap opening and the 
decrease in $\chi$ are closely related 
each other,\cite{takabatake} the microscopic mechanism 
connecting them is not well understood yet.  
At higher temperatures where the gap is not fully developed, 
the $f$- and conduction electrons may form 
local ``Kondo singlets'', resulting in a reduction in $\chi$ 
as observed.\cite{ce343}   On the other hand, the large 
residual $\chi$ at low temperatures is suggestive of a 
Van Vleck-type paramagnetism,\cite{takabatake,footnote} 
which may arise from excited states above the gap.

Another remarkable aspect is that the gap magnitude 
$2\Delta_{opt}$ is much greater than the transport gap 
measured by the dc resistivity, 
$2\Delta_{\rho} \simeq$ 12~meV.\cite{yb12-single}   
In addition, the temperature below which the gap appears 
in $\sigma(\omega)$ is much lower than 
$\Delta_{opt} / k_B \simeq$ 145~K.      These behaviors are 
unexpected for thermal activation of carriers across a 
conventional, direct band gap, since in such a case 
$2\Delta_{opt}$ should be close to $2\Delta_{\rho}$, and 
the gap should appear in $\sigma(\omega)$ 
at $T \sim \Delta_{opt} / k_B$.    
However, if the dispersion near $E_F$ does not form 
a direct gap, $2\Delta_{opt}$ is not necessarily equal 
to $2\Delta_\rho$.   Mutou and Hirashima\cite{mutou} 
performed numerical calculations based on the periodic 
Anderson model in infinite dimensions to analyze the 
temperature-dependent properties of Kondo insulators.  
They showed that $\Delta_{opt} \simeq 2\Delta_\rho$, 
and that the gap appears in $\sigma(\omega)$ 
at $T \simeq (1/2)\Delta_{opt}/k_B$, not at 
$\Delta_{opt}/k_B$.   Similar results were given 
independently by Rozenberg {\it et al.}\cite{rosen}.    
These behaviors arise 
from an {\it indirect} gap formed by the hybridazation 
of narrow $f$- and broad conduction bands renormalized by 
strong correlation effects.   Since $\sigma(\omega)$ mainly probes 
the oscillator strength and the density of states (DOS) 
for direct ($k$-conserving) transitions, 
2$\Delta_{opt}$ is generally larger than 
$2\Delta_{\rho}$ if the gap in the total (both direct and 
indirect) DOS is indirect.    
They also showed that the temperature range for the gap 
development in $\sigma(\omega)$ is nearly the same as 
that for the decrease in $\chi(T)$.   
These theoretical predictions are in good agreement 
with the present experimental results.

These characteristic behaviors of the energy gap 
in $\sigma(\omega)$ and the coincidence of gap 
formation with a decrease 
in $\chi$ are very similar to the results for 
the Ce-based Kondo insulator 
Ce$_3$Bi$_4$Pt$_3$ [Ref.~\onlinecite{ce343}].     
Since these behaviors are shared by the two 
representative Kondo insulators, they are likely to be 
universal optical features of Kondo insulators.   
A difference is seen, however, in the spectral 
shape of the gap in $\sigma(\omega)$: For YbB$_{12}$ 
the spectral weight within the gap is very weak 
at low temperatures, as seen in Fig.~3.    
In contrast, for Ce$_3$Bi$_4$Pt$_3$ the spectral depletion 
within the gap is weaker, with a large ``tail'' 
persisting down to the low-energy 
end of $\sigma(\omega)$.\cite{ce343}

Now we attempt to analyze the low-energy excitations 
above the gap, in particular the shoulder at 40~meV 
and the IR peak at 0.25~eV.    As shown in 
Fig.~4~(a), the IR 
peak can be fitted well by the classical Lorentz 
oscillator model\cite{wooten} 
\begin{equation}
\sigma_L(\omega) \propto 
\frac{\Gamma \omega^2}
{(\omega^2 - \omega_0^2)^2 + \Gamma^2 \omega^2}, 
\end{equation}
where $\omega_0$ is the peak energy and $\Gamma$ is 
the peak width.    In order to separate the shoulder 
and the IR peak, we fitted the IR peak at each 
temperature using (1), then subtracted the fitting function 
from $\sigma(\omega)$.    
The resulting (subtracted) spectra at 78, 50, and 20~K 
are shown in Fig.~4(b).    These spectra reveal that the 
shoulder is actually an asymmetric {\it peak}.   
We attribute this peak to electronic excitations across 
the gap, from mainly Yb 4$f$-derived states 
below $E_F$ to those above $E_F$.   Note that a 4$f$-4$f$ 
transition is optically forbidden, but mixing with other 
symmetry states ({\it e.g.}, Yb 5$d$) may make the transition 
partially allowed.    This interpretation is consistent with 
theoretical calculations\cite{mutou,rosen,harima} and 
photoemission (PE) experiment\cite{susaki}:  Band 
calculations\cite{harima} for YbB$_{12}$ shows a large Yb 
4$f$-derived DOS below and above $E_F$, and numerical 
calculations based on 
the periodic Anderson model also showed large $f$-derived 
DOS below and above the Kondo insulating gap.\cite{mutou,rosen}  
PE experiments\cite{susaki} of YbB$_{12}$ 
found a narrow, asymmetric peak located $\sim$~25~meV 
below $E_F$, which was mainly due to the Yb 4$f$-derived DOS.   
Since $\sigma(\omega)$ probes direct transitions 
between occupied and unoccupied states, we cannot simply 
compare $\sigma(\omega)$ with the PE spectrum, 
which probes the total DOS for the occupied states only.    
Nevertheless, the position of the gap excitation peak in 
$\sigma(\omega)$, 40~meV, is comparable with the peak in 
the PE spectrum, assuming 
that $E_F$ is located near the middle of the gap.

The IR peak is observed for YbB$_{12}$, but not for 
LuB$_{12}$, as shown in Fig.~1.   The most significant difference 
in the electronic states of these two compounds is the position 
of 4$f$-derived states: There is a large Yb 4$f$-derived DOS 
near $E_F$ in YbB$_{12}$ while Lu 4$f$-derived states 
in LuB$_{12}$ are 
located $\sim$ 5~eV below $E_F$.\cite{susaki,harima}   
Therefore, the IR peak is likely to be related to 
Yb 4$f$-derived states near $E_F$.   We tentatively attribute 
the IR peak to optical transitions between the Yb $4f$-derived narrow 
band near $E_F$ and the broad conduction band, which 
mainly consists of B $2p$- and Yb $5d$-derived 
states.\cite{harima}  
Both transitions from the $4f$ band below $E_F$ to 
the $p$-$d$ band above $E_F$ and those from 
$p$-$d$ below $E_F$ to 4$f$ above $E_F$ are possible.   
The much larger integrated strength 
and the much larger width of the IR peak than the gap 
excitation peak 
supports this assignment, since an $f$-$d$ transition is 
optically allowed and the $p$-$d$ bands are 
much broader than the $f$ band.   
Also, the involvement of the Yb $4f$-derived states 
near $E_F$ in 
{\it both the gap excitation peak and the IR peak} 
is consistent with the observation that the spectral 
weight lost in $\sigma(\omega)$ by the gap formation 
is distributed over a wide energy range ($\sim$~1~eV) 
covering the entire IR peak, rather than distributed 
to a narrow energy range directly above the gap.

Anomalous infrared absorption similar to the IR peak 
for YbB$_{12}$ has been also observed for 
rare-earth hexaborides ($R$B$_6$'s) with 
partially-filled 4$f$ shells, including SmB$_6$, CeB$_6$, 
and others.\cite{kimura1}    
The intensity of the observed infrared absorption in 
$R$B$_6$'s with different $R$ elements was found 
to be proportional to the number of 4$f$ electrons, 
and the absorption was attributed to 
an indirect $d$-$d$ transition assisted by exchange 
scattering.\cite{kimura1}   Such specific assignment for 
the IR peak is beyond the scope of the present work, and 
other $R$B$_{12}$'s with different $R$ elements must be 
studied and compared to better understand the nature of the 
IR peak.  

In conclusion, we have reported the first optical reflectivity 
experiment on single crystal YbB$_{12}$.   An energy gap 
formation was clearly observed in $\sigma(\omega)$ below 
$\sim$~70~K.   The gap opening was coincident with a 
large reduction of magnetic susceptibility, showing that the 
gap formation affects the magnetic properties significantly.   
The temperature dependence and the magnitude of the 
gap in $\sigma(\omega)$ showed anomalous behaviors, 
which are probably characteristic of Kondo insulators 
as predicted theoretically.   Two prominent low-energy excitation 
peaks were observed, which were attibuted to 
$f$-$f$ and $f$-$pd$ transitions.

We would like to thank H. Harima for providing unpublished band 
calculations of YbB$_{12}$, LuB$_{12}$, and YB$_{12}$.  
H.O. thanks T. Mutou for many stimulating discussions and for 
providing unpublished simulations on Kondo insulators.  
We acknowlege financial supports from the REIMEI Rsearch 
Resources, the Atomic Energy Research Institute, the 
Electric Technology Research Foundation of Chugoku, and the 
Grants-in-Aid from the Ministry of Education, Science and Culture.

\begin{figure}
\caption{Reflectivity ($R$) and optical 
conductivity ($\sigma$) spectra of YbB$_{12}$ 
and LuB$_{12}$ single crystals at 290~K.}  
\label{fig1}
\end{figure}

\begin{figure}
\caption{Reflectivity ($R$) and 
optical conductivity ($\sigma$) of 
YbB$_{12}$ in the infrared region at 
290~K (dotted-dashed curve), 160~K (dotted), 
78~K (dashed), and 20~K (solid).}  
\label{fig2}
\end{figure} 

\begin{figure}
\caption{Reflectivity ($R$) and optical 
conductivity ($\sigma$) of YbB$_{12}$ at, 
from top to bottom curves for both $R$ and $\sigma$, 
$T$=78, 70, 60, 50, 40, and 20~K.   
The arrow indicates the ``shoulder'' discussed in the text.    
The inset shows the magnetic susceptibility ($\chi$) 
of YbB$_{12}$ single crystal as a function of temperature ($T$).}   
\label{fig3}
\end{figure} 

\begin{figure}
\caption{(a) Optical conductivity of YbB$_{12}$ 
at 78~K (solid curve), a fitting based on (1) (dotted), 
and the resulting spectrum (dashed) obtained after the 
subtraction.   
(b) The resulting spectra obtained after the 
fitting-subtraction procedure at (top to bottom) 
78, 50 and 20~K. }
\label{fig4}
\end{figure} 

\widetext

\begin{references}
\bibitem{fisk} G. Aeppli and Z. Fisk, Comments Cond. Mat. 
Phys. {\bf 16}, 155 (1992).

\bibitem{takabatake} T. Takabatake {\it et al.}, 
J. Magn. Magn. Mat. {\bf 177-181}, 277 (1998). 

\bibitem{wachter1} G. Travaglini and P. Wachter, 
Phys. Rev. B {\bf 29}, 893 (1984).

\bibitem{nanba1} T. Nanba {\it et al}, Physica B 
{\bf 186-188}, 440 (1993).

\bibitem{kimura1} S. Kimura, T. Nanba, S. Kunii and T. Kasuya, 
Phys. Rev. B {\bf 50}, 1406 (1994).

\bibitem{ce343} B. Bucher, Z. Schlesinger, P.C. Canfield 
and Z. Fisk, Phys. Rev. Lett. {\bf 72}, 522 (1994).

\bibitem{yb12-old} M. Kasaya {\it et al.}, J. Magn. Magn. Mat. 
{\bf 31-34}, 438 (1983); {\bf 47{\&}48}, 429 (1985); 
F. Iga, M. Kasaya and T. Kasuya, {\it ibid.}, 
{\bf 52}, 279 (1985); 
{\bf 76 $\&$ 77}, 156 (1988). 

\bibitem{wachter2} P. Wachter and G. Travaglini, 
J. Magn. Magn. Mat. {\bf 47 $\&$ 48}, 423 (1985).

\bibitem{susaki} T. Susaki {\it et al.}, Phys. Rev. Lett. 
{\bf 77} 4269 (1996); Phys. Rev. {\bf B56}, 13727 (1997)

\bibitem{yb12-single} F. Iga, N. Shimizu and T. Takabatake, 
J. Magn. Magn. Mat. {\bf 177-181}, 337 (1998).

\bibitem{bl6a1} M. Sakurai {\it et al.}, 
J. Synchrotron Rad. {\bf 5} 578 (1998).

\bibitem{footnote3} The result of a Hagen-Rubens 
extrapolation was examined by comparing the extrapolated 
$\sigma(\omega)$ with the measured dc conductivity. 

\bibitem{kimura2} S. Kimura {\it et al}., Phys. Rev. B {\bf 46}, 
12196 (1992).

\bibitem{kimura3} S. Kimura {\it et al}.(Physica B, to appear).  

\bibitem{footnote}  The small rise in $\chi(T)$ seen 
for $T \leq 15$~K is due to residual 
impurites\cite{takabatake,yb12-single}.   
For more recently-grown single crystals with less 
impurities, this rise is much weaker, but there still 
remains a large, nearly temperature-independent 
susceptibility, $\chi \sim~3.2 \times 10^{-3}$~emu/mole. 
(F. Iga {\it et al.}, unpublished.)

\bibitem{mutou} T. Mutou and D. Hirashima, J. Phys. Soc. Jpn. 
{\bf 63}, 4475 (1994); {\bf 64}, 4799 (1995).

\bibitem{rosen} M.J. Rozenberg, G. Kotliar, and H. Kajueter, 
Phys. Rev. {\bf B 54}, 8452 (1996).

\bibitem{wooten} F. Wooten, {\it Optical Properties 
of Solids} (Academic Press, New York, 1972).

\bibitem{harima}  H. Harima, A. Yanase and T. Kasuya, 
J. Magn. Magn. Mat. {\bf 47\&48}, 567 (1985); 
A. Yanase and H. Harima, Prog. Theor. Phys. Suppl. 
{\bf 108}, 19 (1992); H. Harima, unpublished (1998).

\end{references}
\end{document}